\documentclass[13pt]{elsart}
\usepackage{graphics,epsfig,graphicx,amssymb,latexsym}
\journal{Computers \& Graphics}

\date{27 November 2002}

\begin{document}


\begin{frontmatter}
\title{Application of interactive parallel visualization for
       commodity-based clusters using visualization APIs}
\author{Stanimire Tomov\corauthref{cor}},
\corauth[cor]{Corresponding author.}
\ead{tomov@bnl.gov}
\author{Robert Bennett},
\ead{robertb@bnl.gov} 
\author{Michael McGuigan},
\ead{mcguigan@bnl.gov}
\author{Arnold Peskin},
\ead{peskin@bnl.gov}
\author{Gordon Smith},
\ead{smith3@bnl.gov}
\author{John Spiletic}
\ead{spiletic@bnl.gov}
\address{Information Technology Division, 
         Brookhaven National Laboratory, 
         Bldg. 515, Upton, NY 11973}


\begin{abstract}      
We present an efficient and inexpensive to develop application for
interactive high-performance parallel visualization.
We extend popular APIs such as
Open Inventor and VTK to support commodity-based cluster visualization.
Our implementation follows a standard master/slave concept: the general
idea is to have a ``Master'' node, which will intercept a sequential
graphical user interface (GUI) and broadcast it to the ``Slave''
nodes. The interactions between the nodes are implemented 
using MPI. The parallel remote rendering uses 
Chromium.
This paper is mainly the report of our implementation experiences.
We present in detail the proposed model and key aspects of its implementation. 
Also, we present performance measurements, we benchmark 
and quantitatively demonstrate the dependence of the visualization speed
on the data size and the network bandwidth, and we identify the singularities
and draw conclusions on Chromium's sort-first rendering architecture.
The most original part of this work is the combined use of Open Inventor 
and Chromium.
\end{abstract}



\end{frontmatter}


\section{Introduction}

\begin{figure}[ht]
\centerline{
  \includegraphics[width=1.7in]{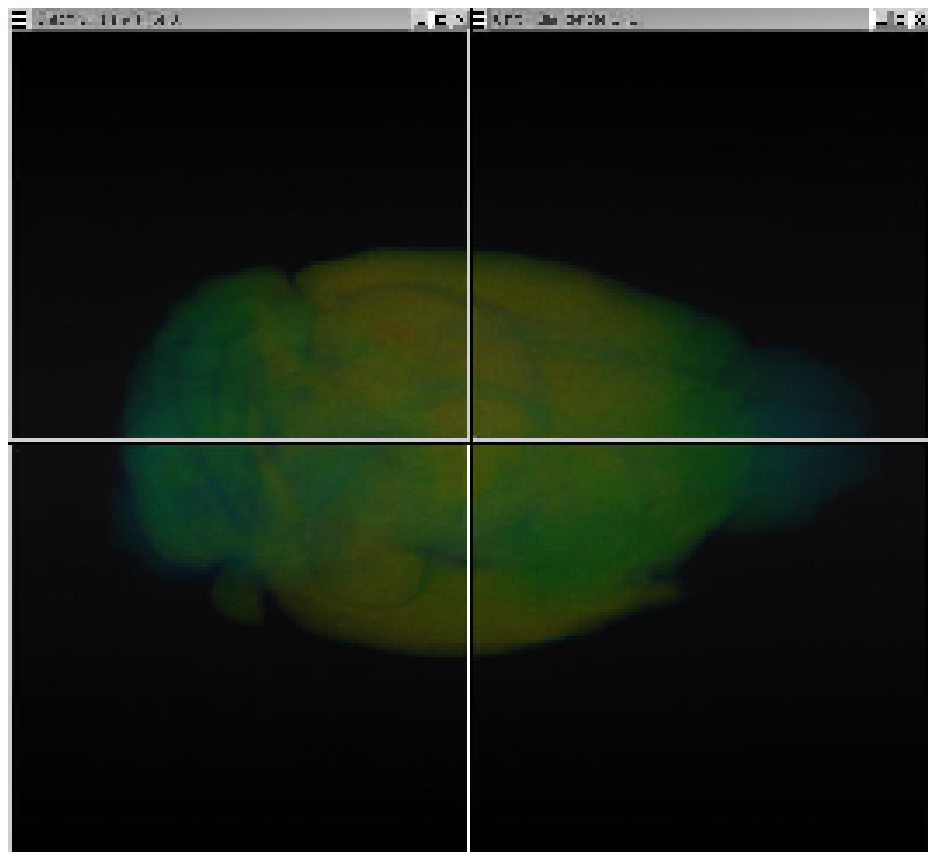}\hspace{0.1cm}
  \includegraphics[width=1.92in]{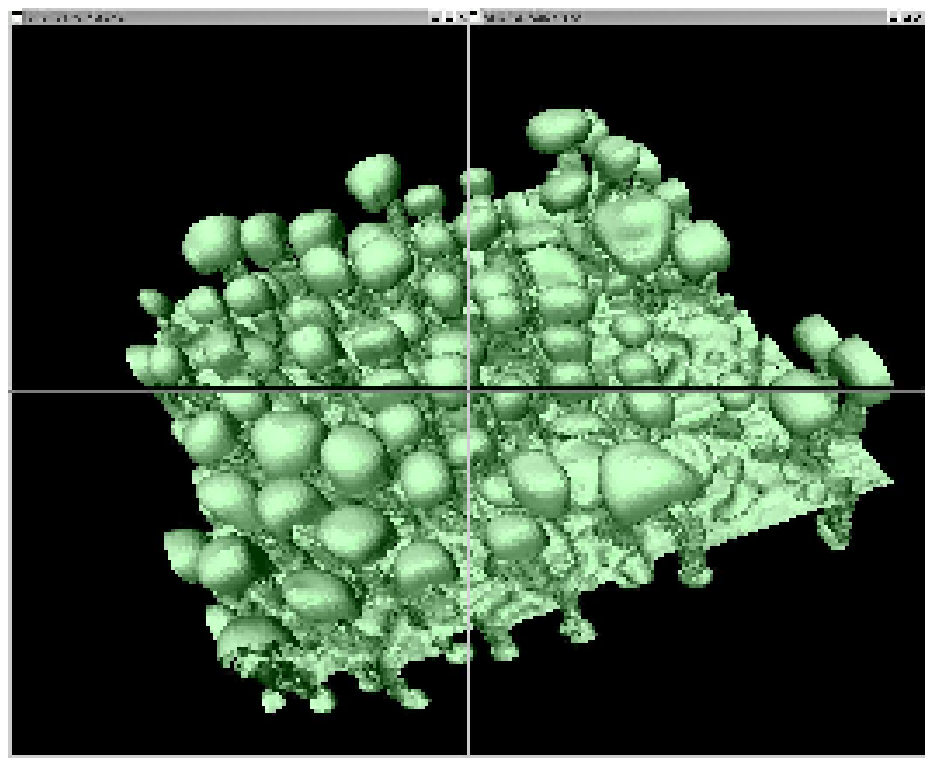}\hspace{0.1cm}
  \includegraphics[width=1.7in]{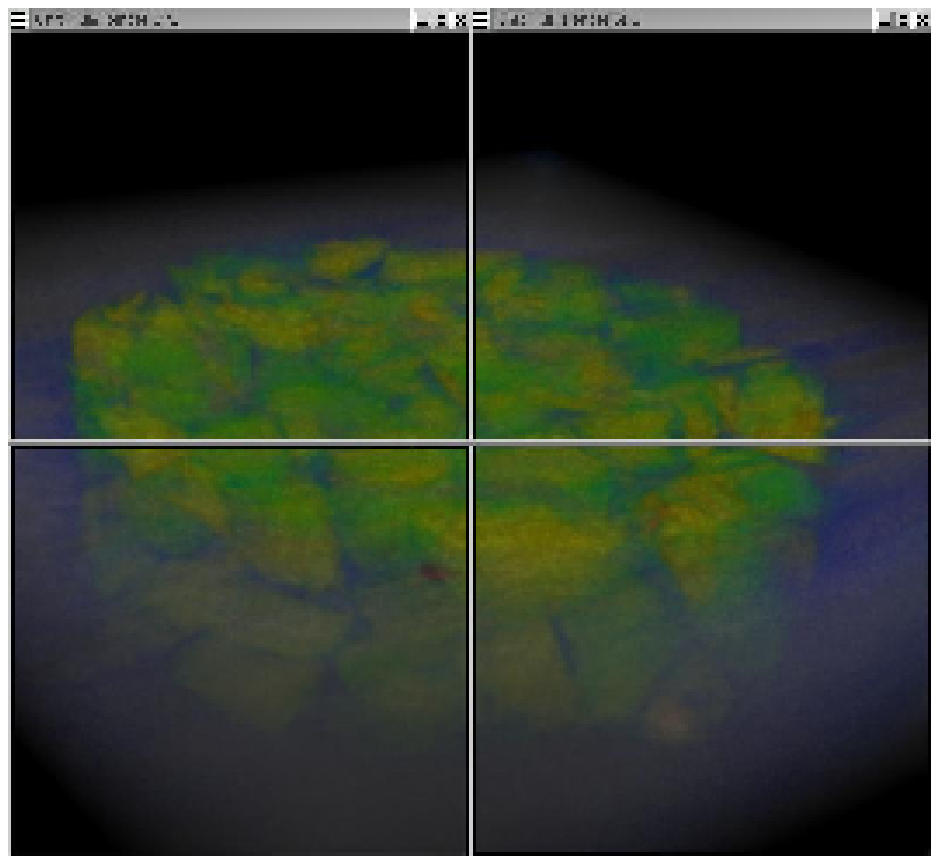}
           }
\caption{Left: mouse brain volume visualization on a
         $512$ X $256$ X $256$ grid;
         Middle: Fluid flow isosurface visualization;
         Right: Material micro-geometry studies: crushed rock
         volume visualization on a $1007$ X $1007$ X $256$ grid.
         The three examples are rendered
         in parallel on a display composed of $4$ tiles.}
\label{applications}
\end{figure}

We are interested in large scale high-performance 
scientific parallel visualization with 
real-time interaction. More precisely, we want to support the interactive
extraction of insight from large sets of data. Also,
in order not to miss important details in the data, we 
would like to develop high resolution visualization.
Our interest is 
motivated by today's technological advances, which stimulate and 
facilitate the development of increasingly complicated mathematical
models. Visualization with real-time interaction is essential for 
better analysis and understanding of the masses of data produced by 
such models, or by devices such as Magnetic Resonance Imaging (MRI)
and laser-microscope scanners. 
Some of the applications that we are interested in and work on are
given on Figure \ref{applications}. Others come from high-energy
physics, climate modeling, etc. Currently the size of the data sets 
that we are dealing with is around $1$ GB. 

We develop an inexpensive and, as the results show, efficient application.
The approach  is as follows:
\begin{itemize}
  \item Do remote parallel rendering to a large tiled display
        (see for example Figure \ref{applications} where we show displays
         composed of $4$ tiles).
  \item Use commodity-based clusters connected with high speed network.
  \item Extend and combine already existing APIs such as Chromium,
        Open Inventor, VTK, etc.
\end{itemize}
The APIs considered are open source. Open Inventor
(see \cite{OpenInventor,OIreference}) is a library of 
C++ objects and methods for 
building interactive 3D graphics applications,
VTK (see \cite{vtk,vtkuserguide}) is another widely used 
API for visualization and image processing, and
Chromium (see \cite{Chromium,Ch1,Ch2,Ch3}) is an OpenGL 
\cite{opengl} interface for cluster visualization
to a tiled display. Chromium is based on WireGL \cite{wiregl}
and provides a scalable display technology.
The latest Chromium paper \cite{Ch4} reports on the implementation 
of components (stream processing units, or SPUs) that can be used
in sort-last parallel graphics applications. More on the sort-last 
parallel rendering approach can be found in \cite{magallon} and \cite{wylie}.

There are feasible alternatives to our approach. For example,
the tiled display can be replaced with a high resolution flat 
panel driven by a single workstation, the
visualization clusters can be replaced with fast sequential machines,
and the process of 
extending and combining already existing APIs can be replaced with
building the entire visualization system from scratch.
The IBM's T221 display with its maximum resolution of 
$3840 \times 2400$ brings to the scientist's desktop a resolution
that is high enough for many applications. For such cases this is a
preferred alternative and we have it as an option in our implementation.
Similarly, for ``small'' data sets we prefer sequential processing.
Extending and combining visualization APIs is
a popular and in many cases preferred development 
approach since one can easily leverage
already existing and powerful APIs. Examples are the 
extension of VTK to ParaView (see \cite{vtkpublic}) and parallel 
VTK (see \cite{paravtk}), {\tt VisIt} (see \cite{visit}), etc.
Building an entire visualization system from scratch
may be efficient for very specific requirements, but
in general is expensive and time consuming (see for example 
NASA's long term project {\tt ParVox} \cite{nasa}).
Our goals, apart from developing the application outlined above,
also include studying and identifying the singularities of 
the Chromium API. We concentrate on Chromium's sort-first rendering,
a model inherited, and further developed from the WireGL API.

The parallel model that we consider is MIMD (multiple instructions --
multiple data). MIMD parallel models are common in the design of parallel 
visualization software for clusters of workstations. A fundamental
framework is when cluster nodes process separate parts of a global scene and
their output is composited and rendered to a tiled display.
Providing MIMD model visualization systems 
with efficient user interaction has become a task of great interest.
See for example \cite{workshop}. 
The continuous interest also prompted the development of
a new cluster rendering utility toolkit (CRUT) which was reported 
in \cite{CRUT}. CRUT is a glu-like toolkit that will facilitate 
the development of user interaction within Chromium, and hence 
the development of applications and visualization API extensions
like the one that we developed.
We pursue a Master/Slave paradigm: we declare one of the
cluster nodes as GUI Master, use the Master to intercept the 
sequential user input, and broadcast that input in a user defined protocol
to the other nodes (here called Slave nodes). This yields a system 
where the GUI Master sends to the Slave nodes instructions of ``how and when'' 
to redraw their part of the scene. The user interface in this case
is part of the visualization software. Another implementation is
to have the GUI Master separate from the visualization software.
The interaction in this case would be through a GUI window that will 
simulate a ``parallel interaction device'' that takes a sequential 
user input and broadcasts it to the cluster nodes (through sockets).
The details are given in the following sections.

The article is organized as follows. In Section \ref{model} we describe 
our model framework and its implementation in extending Open Inventor and VTK. 
Next (in Section \ref{results}) we discuss issues that are 
important for the development of high performance visualization on clusters
of workstations. Also, we provide some of our
performance results. The last section (Section \ref{conclusions})
summarizes the results of this work.

\section{Parallel GUI model and implementation}\label{model}

\begin{figure}[ht]
\centerline{
  \includegraphics[width=4.6in]{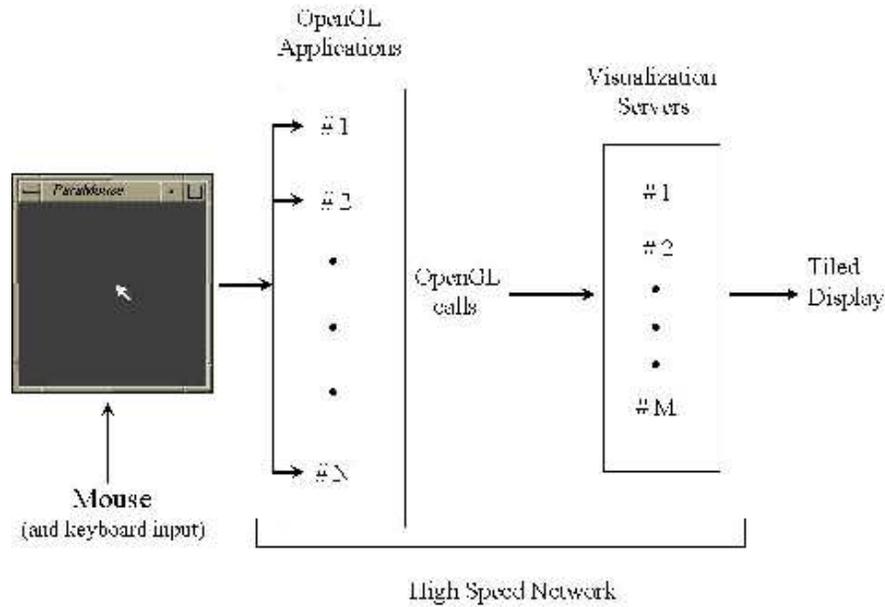}
           }
\caption{Visualization pipeline. The parallel GUI (ParaMouse) gets the 
         sequential user input (through mouse, keyboard, etc.) and broadcast 
         it to the OpenGL applications through sockets. The applications 
         use Chromium to render to a large tiled display.}
\label{pipeline}
\end{figure}

Parallel GUI for a MIMD parallel visualization model requires the presence
of a Master node that will synchronize with the other nodes the redrawing of
the consecutive scenes according to a sequential user input. 
We consider two variations of parallel GUI.
Both have similar visualization pipelines (see Figure \ref{pipeline}).
The parallel GUI (ParaMouse) gets the sequential user input (through mouse, 
keyboard, etc.) and broadcasts it to the OpenGL applications through sockets. 
Every application is visualizing a separate part of a global scene. The 
applications are bound together by MPI communications. The OpenGL 
calls that the applications make are intercepted by Chromium and sent
through the network to the visualization servers, which composite the input 
and render it to a tiled display.

Chromium's parallelization model (denoted in the article by CPM) is 
represented with the pseudo-code:
\begin{verbatim}

  glXMakeCurrent(getDisplay(), getNormalWindow(), 
                 getNormalContext());
  if (clearFlag)
    glClear(GL_COLOR_BUFFER_BIT | GL_DEPTH_BUFFER_BIT);
  glBarrierExecCR(MASTER_BARRIER);

  do sequential OpenGL rendering of the local scene

  glBarrierExecCR(MASTER_BARRIER);
  if (swapFlag)
    glXSwapBuffers(getDisplay(), getNormalWindow());
  else
    glXSwapBuffers(getDisplay(), CR_SUPPRESS_SWAP_BIT);

\end{verbatim}
where the display obtained is {\bf composited} if 
{\tt clearFlag} is $1$ for all processors, {\tt swapFlag} is $1$ for processor 
with rank $0$, and $0$ for the rest, and {\bf tiled} if {\tt clearFlag} 
and {\tt swapFlag} are $1$ only for processor with rank $0$.

To extend Open Inventor and VTK (or any API) to support interactive 
cluster visualization within the above framework we did the following:
\begin{itemize}
  \item Apply the CPM to the APIs rendering method(s).
  \item Implement the ParaMouse parallel GUI or extend the API's
        GUI using the Master-Slave concept.
\end{itemize}
To implement the first step for Open Inventor we extended the 
SoXtRenderArea::redraw() method. For VTK this step is
implemented by David Thompson, Sandia National Laboratory \cite{vtkChromium}.
We implemented in Open Inventor the Master-Slave concept
by extending the GUI that {\tt ivview} provides.
Function main in {\tt ivview}
was extended by implementing the pseudo-code:
\begin{verbatim}

  declare processor with rank 0 as Master;
  if (Master)
    run GUI as implemented in ivview;
  else
    listen for instructions from the Master;

\end{verbatim}
We extended the callback functions
triggered by the devices that Open Inventor's GUI handles
(mouse, keyboard, etc). The callback functions responding to
user input have {\tt MPI\_Isend}  
of the invoked user input in the Master node
to the Slave nodes.
The Slave nodes are in ``listening'' mode ({\tt MPI\_Recv} or
{\tt MPI\_Irecv} while animating), which is given with the
pseudo-code below, and upon receiving data,
representing instructions in an internally defined protocol, they
call the action for the invoked input and redraw their part 
of the global scene (if necessary). 
\begin{verbatim}

  // Initialize Inventor
  SoDB::init();
  SoNodeKit::init();
  SoInteraction::init(); 
    
  // Build the Inventor's objects and scene graphs 
  SoSeparator *root = new SoSeparator;
  root->ref();    
  readScene(root, files);
    
  // Create a Slave node ExaminerViewer
  SoExaminerViewer *viewer = new SoExaminerViewer(crrank);
  viewer->setSceneGraph(root);

  // Chromium initialization    
  crctx = crCreateContextCR(0x0, visual);
  crMakeCurrentCR(crwindow, crctx);
  glBarrierCreateCR( MASTER_BARRIER, crsize);

  glEnable(GL_DEPTH_TEST);
  viewer->mainLoop();

\end{verbatim}
The SoExaminerViewer class is based on the Open Inventor's 
SoXtExaminerViewer class. The difference is that 
SoExaminerViewer does not have Xt window interface function
calls and the rendering is with the CPM. 

\begin{figure}[ht]
\centerline{
  \includegraphics[width=2.35in]{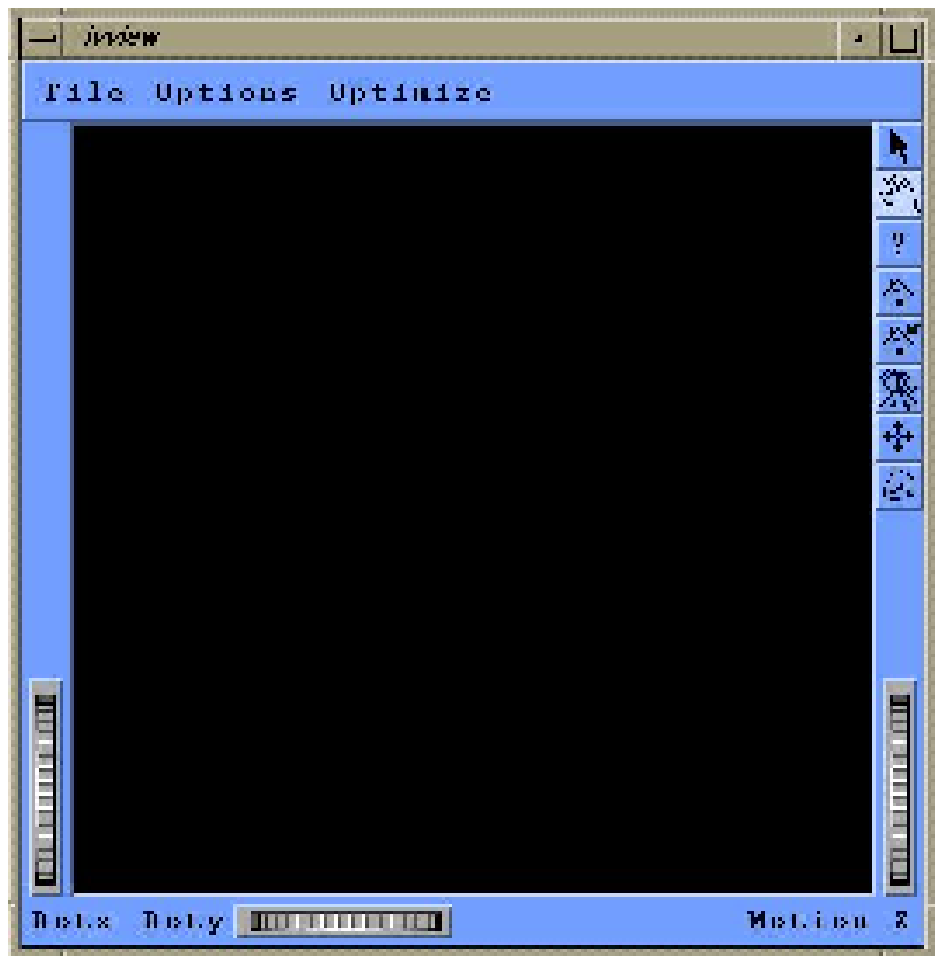}\hspace{0.5cm}
  \includegraphics[width=2.3in]{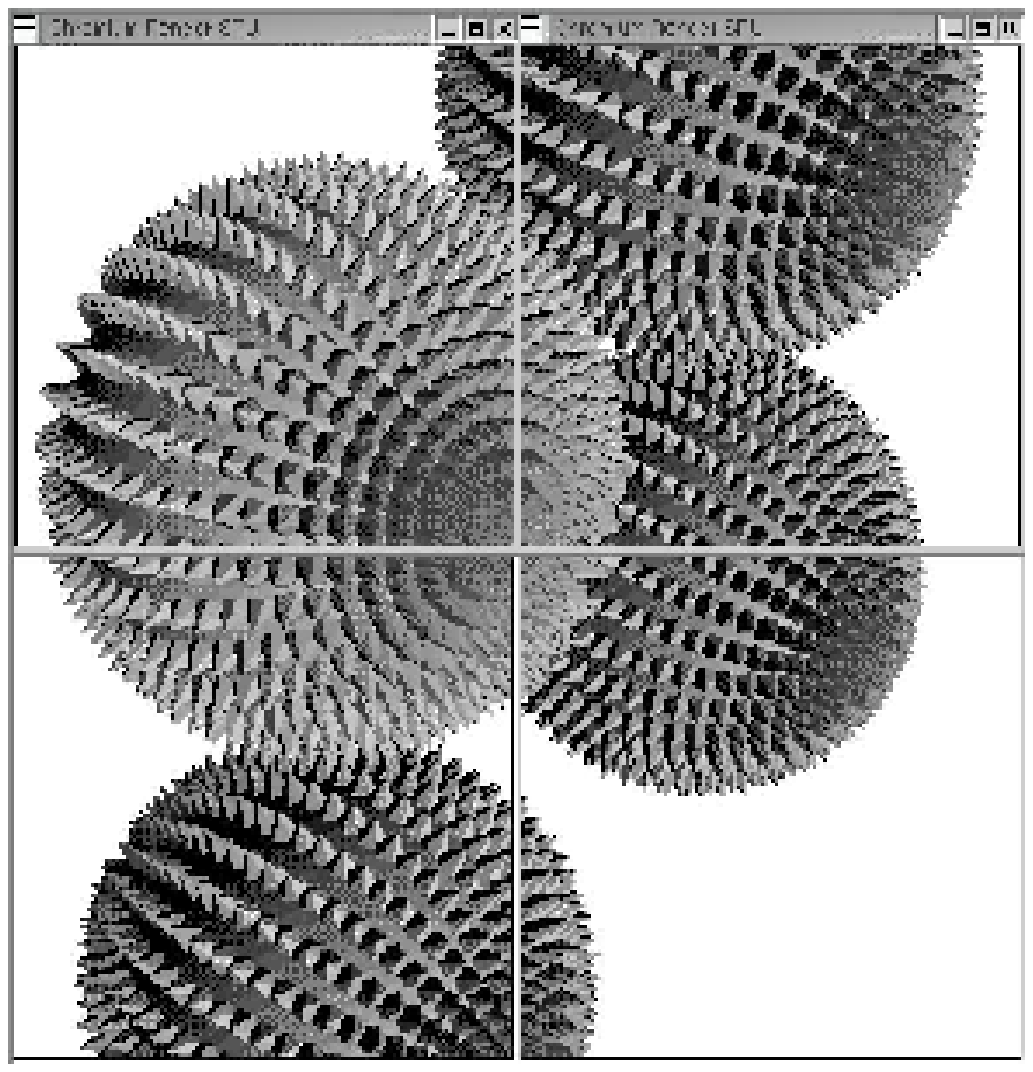}
           }
\caption{Left: {\tt ivview} window interface;
         Right: example of a display with 4 tiles.\newline
         The scene is rendered
         in parallel on the tiled display. The interactive
         user interface is through the {\tt ivview} window, with the 
         functionality that {\tt ivview} provides.}
\label{interface}
\end{figure}

The user interface is through the {\tt ivview} window (see Figure 
\ref{interface}, left), which is blank and used only for the user
input. The scene is drawn in separate windows/tiles.
The example from Figure \ref{interface} (right) shows a display with $4$
tiles. There are $4$ applications running, each of which visualize
a sphere. 

In VTK's Master-Slave model we extend the 
vtkXRenderWindowInteractor class by:
\begin{itemize}
  \item Implementing ``pure'' X Windows GUI (no GL calls).
  \item Adding MPI send/receive calls for mouse and 
        keyboard events.
\end{itemize}
 
For the ParaMouse interface we added in VTK a new interactor style,\\
called {\tt vtkInteractorStylePMouse}.
 
\section{Observations and performance results}\label{results}
We did our implementation and testing on a Beowulf cluster with $4$ nodes,
each node with $2$ Pentium III processors, running at 1 GHz.
Every node has {\tt Quadro2 Pro} graphics card.
The nodes are connected into a local area network with communications running
through $100$ Mbit/sec fast Ethernet or $1$ Gbit/sec fiber optic network.
More about the general performance of this particular machine can be found in 
\cite{vito}.

The experience that we had in developing interactive parallel visualization
is summarized as follows:
\begin{enumerate}
   \item For the parallel model considered performance is problem
         and interaction specific. \\
         For example, depending on the data and the user interaction, 
         the entire global scene may be mapped to a single tile 
         of the display. Also, the Chromium bucketing strategy 
         \cite{Ch1,Ch2,Ch3} will not work for scenes composited of 
         non-localized consecutive polygons.
   \item The GUI communications time is negligible compared to the 
         visualization time, which leads us to the scalability results 
         reported in \cite{Ch1,Ch2,Ch3}. This statement is supported by 
         our next observation.
   \item We can send approximately $5,000$ ``small'' (less then 100 bytes) 
         messages/sec (see \cite{vito}).
   \item Chromium automatically minimizes data flow, except for geometry
         flow (see below).
   \item It is advisable to keep small static scenes in display 
         lists (see below).
\end{enumerate}

The network data flow is usually the bottleneck in the visualization pipeline
that we consider. Chromium provides several techniques to minimize it. 
They are: simplified network protocol, bucketing, and state tracking
(see \cite{Ch1,Ch2,Ch3}). None of these techniques however are intended 
for the automatic minimization of the geometry flow, which usually is the 
most expensive component. For example in animation the same objects are drawn 
without (or with minimal) change from frame to frame. Nevertheless the 
scene is transmitted over the network for every frame, creating an enormous
bottleneck, as shown in the next example. 
Data flow minimization techniques that exploit spacial or temporal coherence
of consecutive frames (see for example \cite{yoon}) are not supported 
within the current framework.

The following example illustrates items $4$ and $5$ from the
above observations. We use a small static scene 
composed of spheres with spikes, as the ones shown on Figure 
\ref{interface}, right. Each sphere is composed of $11,540$ triangles. 
Sequentially, one sphere is drawn by the Open Inventor at a rate of $195$ 
frames/sec, i.e. $2,251,569$ triangles/sec. For $2$ spheres the rate is 
$117$ frames/sec or $2,709,130$ triangles/sec, etc. 
The speed of visualizing $2$ spheres on $2$ processors ($2$ tiles)
is approximately $6$ frames/sec. The composited visualization
($2$ processors, $1$ tile) is approximately $3$ frames/sec.
Runs with various problem sizes and parallel visualization configurations
give similar results in favor of the sequential execution. 

The enormous difference is due to the fact that in the first case
the scene resides in the graphics card memory, while in the second
the same scene is transmitted over the network for every frame.
Table $1$ gives more results related to the network's
bandwidth bottleneck and the overall performance of the system. 
The results are for different applications using
the $100$ Mbit and the $1$ Gbit network. The first one, mouse brain,
comes from medical science and is a 3D surface of a mouse brain.
The size of the data is $16$ MB. The $100$ Mbit network gets
saturated and the frame rate of $0.46$ frames/second is expected,
since the data traffic, although dependent on user interaction and
locality of the scene polygons, is proportional to the data size.
The $0.46$ frames/second translates to $2.17$ seconds per frame.
The network transfer takes $1.75$ seconds. 
The next application, fluid flow,
is an isosurface extracted from a fluid flow simulation data. The beetle head
is also an isosurface. It was extracted from the X-ray computed microtomography
data of an Alaskan spruce bark beetle.
With performance mainly depending on the data size 
we observe that doubling the data size reduces the frame rate two times.
Another application type that we tested is ray tracing volume visualization.
We applied it  
to X-ray computed microtomography data of a rock sample of size $1$ GB. 
This is an 
example where a substantial part of the visualization is done
in the CPU and only the result, in terms of
OpenGL primitives, is sent through the network.
On the $100$ Mbit
network we get $12.5$ seconds per frame. $10.1$ seconds are spent in the
ray tracing algorithm (run on $4$ cluster nodes with dual 
processor on each node) and $2.4$ seconds in transfer and rendering of 
the OpenGL
commands issued in the ray tracing algorithm. 

Switching to the $1$ Gbit network approximately doubles the performance
results. Similar improvement was observed
in \cite{vito} for various numerical analysis applications.
The ray casting relies mostly on the CPU's performance
and switching to the $1$ Bbit network did not show any speed up.

\vspace{0.2in}
\begin{table}[ht]
\begin{center}
\begin{tabular}{|c|c|c|c|} \hline
  &   & \multicolumn{2}{|c|}{Frames per second using}\\
\cline{3-4}
 \raisebox{2.0ex}[0pt]{Application}            &  
 \raisebox{2.0ex}[0pt]{Size}      &  $100$ Mbit network & $1$ Gbit network   \\ \hline 
 mouse brain & 16 MB  &  $0.46$             &    $0.87$          \\ \hline
 fluid flow  & 40 MB  &  $0.20$             &    $0.53$          \\ \hline
 beetle head & 80 MB  &  $0.09$             &    $0.25$          \\ \hline
\end{tabular}
\end{center}
\vspace{0.1in}
{\bf Table $1$.}~ Dependence of the visualization speed
                    on the data size and the network
                  bandwidth for different applications. The 
                    global scene is split into $4$ 
                  and rendered by $4$ Chromium rendering SPUs 
                    to $4$ tiles display.
\label{TABLE1}
\end{table}
\vspace{0.1in} 

\begin{figure}[ht]
\centerline{
  \includegraphics[width=1.8in]{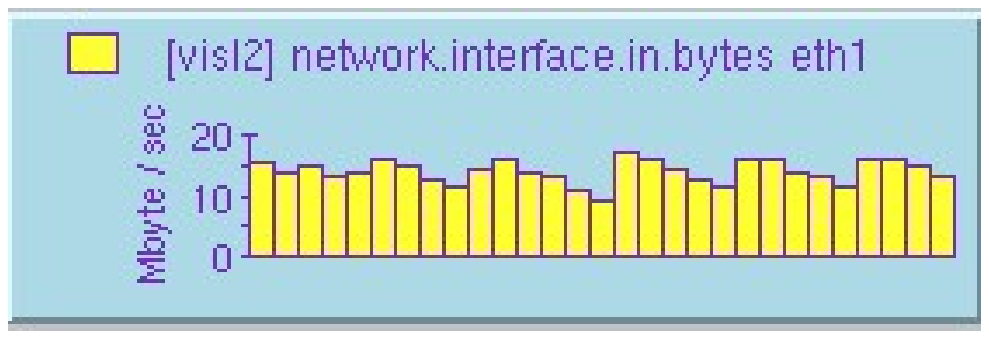}\hspace{0.1cm}
  \includegraphics[width=1.8in]{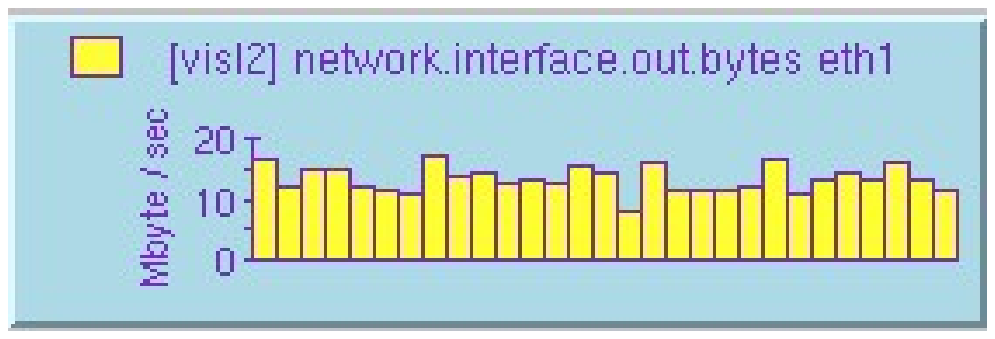}\hspace{0.1cm}
  \includegraphics[width=1.8in]{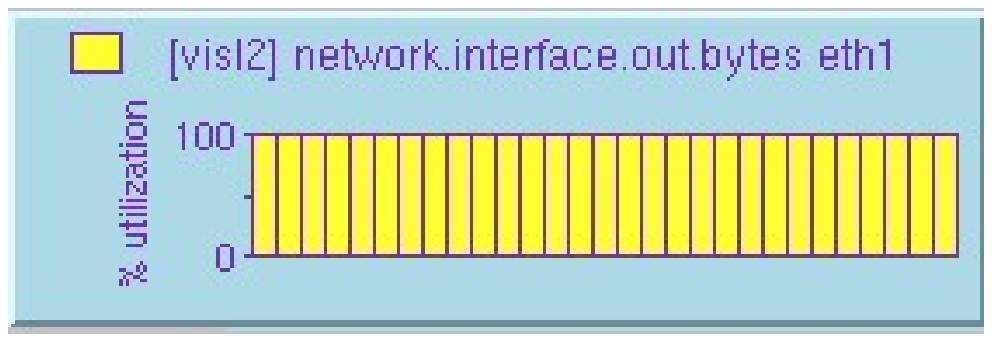}
           }
\caption{Network utilization for the mouse brain application (data size 
         $16$ MB). We show the traffic for one of the $4$ cluster nodes
         over the $1$ Gbit network.
         Left: {\tt traffic in} in Bytes;
         Middle: {\tt traffic out} in Bytes;
         Right: network utilization.}
\label{network}
\end{figure}

We used SGI's {\tt pmchart} to monitor the network traffic for the different
applications described above and in Table $1$. 
The network gets saturated for the applications discussed. 
The network traffic {\tt in} and {\tt out} of every cluster node looks like
the one given on Figure \ref{network}.

\vspace{0.2in}

The non-automatic mechanism that Chromium provides for 
acceleration (minimization)
of the geometry flow is display lists. They are supported 
by sending the lists to each rendering server,
and thus guaranteeing their presence on the server when they are called.
Display lists in Open Inventor are  created for the parts
of the scene that have as root
{\tt SoSeparator} nodes with field {\tt renderCaching} turned
{\tt ON} or {\tt AUTO} (for more information see \cite{OpenInventor}, 
pages $224$-$227$). Using display lists we get a speed of $90$ frames/sec
for visualizing $2$ spheres by $2$ processors on composited display, and 
$80$ frames/sec on tiled display. Different parallel and display
configurations illustrate the same order of improvement. These results 
are comparable to the sequential ones. 
The trade-offs are that the process is not automatic and that the display 
lists are broadcast to all rendering servers
and thus replicating the scene in each node. Work on the automatic 
optimization of the geometry data flow can be found in \cite{stream_caching}.

\section{Conclusions}\label{conclusions}
We developed an application for interactive parallel visualization on 
tiled displays for commodity-based clusters. We used Chromium's parallel
rendering and extended popular visualization APIs, such as Open Inventor
and VTK. The most original part of this work is the combined use of
Open Inventor and Chromium. We gave implementation details and described 
our experience in the combined use of the Chromium's rendering technology 
with Open Inventor and VTK.
A general conclusion is that fast sequential visualization often relies on
graphics acceleration hardware, data sampling methods, static scenes, and
data size within the hardware limitations, while the visualization developed
does not require expensive graphics hardware, provides high resolution,
facilitates well the visualization 
of time-varying scenes, and can handle large data sets.
We demonstrated a low cost (time, money, effort, etc.) of development. 
Our benchmarks quantitatively demonstrated the bottleneck that
the network's bandwidth presents in the sort-first rendering architecture.
This bottleneck makes the sort-first architecture more appealing for
time-varying data sets, commodity-based clusters with slow (or no) 
graphics acceleration, and visualization algorithms that rely on the CPU's 
speed, such as ray casting.

\vspace{0.1in}
{\bf {Acknowledgments}}.~
We would like to thank Beverly Tomov from Cold Spring Harbor Laboratory, NY, 
for her attentive editing and remarks. 
We thank the anonymous referees whose constructive remarks
improved the presentation. 
We also thank Keith Jones from the BNL's Environmental Research and 
Technology Division for providing the data produced by the
synchrotron-based computed microtomography at the Brookhaven 
National Synchrotron Light Source.



\begin{thebibliography}{99}

\bibitem{paravtk}
  J.Ahrens, C.Law, K.Martin, M.Papka, W.Schroeder,
  {\em A Parallel Approach for Efficiently Visualizing Extremely
    Large, Time-Varying Data Sets},
   Technical Report \#LAUR-00-1620 (Los Alamos National 
  Laboratory, NM, USA, 2000).

\bibitem{CRUT}
  Dale Beermann,
  {\em CRUT: The Cluster Rendering Utility for Chromium},
  Internet address (accessed on 05/2003):\\
  http://www.cs.virginia.edu/{\tiny $\!\!\ ^\sim$}drb3t/crut.ps

\bibitem{visit}
  E.Brugger,
  {\em Visit: A Component-Based, Parallel Visualization Package},
  DOE, Computer Graphics Forum 2001,
  Internet address (accessed on 08/2002): \\
  http://www.emsl.pnl.gov:2080/docs/doecgf2001/abstractlist.html

\bibitem{Ch2}
  I.Buck, G.Humphreys, and P.Hanrahan, 
  {\em Tracking Graphics State for Networked Rendering},
  Proceedings of SIGGRAPH/Eurographics Workshop on Graphics Hardware
  (Interlaken, Switzerland, 2000) 87-95.

\bibitem{stream_caching}
  Nathaniel Duca, Peter Kirchner, James Klosowski,
  {\em Stream Caching: Optimizing Data Flow within Commodity Visualization
       Clusters}, Workshop on Commodity-Based Visualization Clusters, October 2002.

\bibitem{MPI}
  W. Gropp,
  {\em  Tutorial on MPI: The Message-Passing Interface},
  Internet address (accessed on 06/2002): \\
  http://www-unix.mcs.anl.gov/mpi/tutorial/index.html.

\bibitem{Chromium}
G. Humphreys, 
{\em Chromium Documentation, Version BETA},
Internet address (accessed on 06/2002): \\
http://chromium.sourceforge.net/

\bibitem{Ch1}
  Greg Humphreys, Ian Buck, Matthew Eldridge, Pat Hanrahan,
  {\em Distributed Rendering for Scalable Displays},
  Proceedings of Supercomputing 2000 (Dallas, TX, USA, 2000).

\bibitem{wiregl}
  G. Humphreys, M. Eldridge, I. Buck, G. Stroll, M. Everett, P. Hanrahan,
  {\em WireGL: A Scalable Graphics System for Clusters},
  Proceedings of SIGGRAPH 2001 (Los Angeles, CA, USA) 129-140.

\bibitem{Ch4}
  Greg Humphreys, Mike Houston, Ren Ng, Randall Frank, Sean Ahern, 
  Peter D. Kirchner, James T. Klosowski,
  {\em Chromium: A Stream-Processing Framework for Interactive Rendering
       on Clusters},
  Proceedings of SIGGRAPH 2002 (San Antonio, TX, USA) 693-712.

\bibitem{Ch3}
  Homan Igely, Gordon Stoll, and Pat Hanrahan,
  {\em The Design of a Parallel Graphics Interface},
  Proceedings of SIGGRAPH 1998 (Orlando, FL, USA) 141-150.

\bibitem{vtkpublic}
{\em ParaView: Parallel Visualization Application},
Internet address (accessed on 08/2002): \\
http://www.paraview.org/HTML/Index.html

\bibitem{magallon}
Marcelo Magallon, Matthias Hopf, Thomas Ertl,
  {\em Parallel Volume Rendering Using PC Graphics Hardware},
  Proceedings of the 2001 Pacific Graphics Conference.

\bibitem{vito}
V. Mirinavicius,
  {\em Investigation of MPI performance in Parallel Linear Algebra
       Software on Linux Beowulf Supercomputers},
  Technical Report (to appear),
  Brookhaven National Laboratory, Upton, NY, 2002.

\bibitem{nasa}
NASA, Jet Propulsion Laboratory,
{\em ParVox, A Parallel Splatting Volume Rendering System},
Internet address (accessed on 08/2002): \\
http://alphabits.jpl.nasa.gov/ParVox/

\bibitem{opengl}
  { J. Neider, T. Davis, and M. Woo} (OpenGL Architecture Review Board),
  {\em OpenGL Programming Guide : The Official Guide to Learning 
       OpenGL, Release $1$},
  Addison-Wesley Publishing Company, Boston, May 1996.

\bibitem{vtk}
 W. Schroeder, K.Martin, L.Avila, C. Law 
 {\em The Visualization Toolkit User's Guide},
 Kitware, Inc.

\bibitem{vtkuserguide}
 W. Schroeder, K.Martin, B.Lorensen,
 {\em The Visualization Toolkit: An Object-Oriented Approach to $3$D Graphics},
 Prentice-Hall PTR.

\bibitem{vtkChromium}
 David Thompson and Gary Templet,
 {\em CPlant, VizClusters, VTK, Chromium},
 DOE Computer Graphics Forum - April, 2002 (Montauk, NY), 
 Internet address (accessed on 08/2002): \\
 http://www.ccd.bnl.gov/visualization/doecgf\_2002/

\bibitem{workshop}
 IEEE Visualization 2002, 
 workshop on: {\em Commodity-Based Visualization Clusters},
 Proceedings of IEEE Visualization, 2002.

\bibitem{OpenInventor}
J. Wernecke, 
{\em The Inventor Mentor : Programming Object-Oriented 3D Graphics with 
     Open Inventor}, 
Addison-Wesley Publishing Company, NY, November 1995.

\bibitem{OIreference}
{\em Open Inventor C++ Reference Manual: The Official Reference Document for Open Inventor, Release 2},
Addison-Wesley Publishing Company, January 1994. 

\bibitem{wylie}
  Brian Wylie, Constantine Pavlakos, Vasily Lewis, and Ken Moreland,
  {\em Scalable Rendering on PC Clusters},
  IEEE Computer Graphics and Applications, July/August 2001 (Vol.21, No.4), pp. 62-70.

\bibitem{yoon}
  Ilmi Yoon and Ulrich Neumann,
  {\em Web-Based Remote Rendering with IBRAC},
  Euroghraphics 2000 (Volume 19, Number 3).

\end{thebibliography}
\end{document}